\documentclass[preprint,superscriptaddress, nofootinbib]{revtex4}
\usepackage{graphicx}% Include figure files
\usepackage{dcolumn}% Align table columns on decimal point
\usepackage{slashbox,multirow}
\usepackage{amsmath,amsthm,amssymb}
\begin{document}
%

%\preprint{APS/123-QED}

\title{A note on late-time tails of spherical nonlinear waves}

\author{Piotr Bizo\'n}
\affiliation{M. Smoluchowski Institute of Physics, Jagiellonian
University, Krak\'ow, Poland}
\author{Tadeusz Chmaj}
\affiliation{H. Niewodniczanski Institute of Nuclear
   Physics, Polish Academy of Sciences,  Krak\'ow, Poland}
   \affiliation{Cracow University of Technology, Krak\'ow,
    Poland}
\author{Andrzej Rostworowski}
\affiliation{M. Smoluchowski Institute of Physics, Jagiellonian
University, Krak\'ow, Poland}

\date{\today}
\begin{abstract}
We consider the long-time behavior of small amplitude solutions of the semilinear wave equation
$\Box \phi =\phi^p$ in odd $d\geq 5$ spatial dimensions. We show that for the quadratic
nonlinearity ($p=2$) the tail has  an anomalously small amplitude and fast decay. The extension
of the results to more general nonlinearities involving first derivatives is also discussed.
\end{abstract}

\maketitle

In a recent paper \cite{bcr} we studied the late-time tails of spherical waves propagating on
even-dimensional Minkowski spacetime under the influence of a long range radial potential. Using
 perturbation methods we showed that in six and higher even dimensions there exist exceptional
potentials which produce tails with anomalously small amplitudes and fast decay rates. The main
purpose of this note is to show that anomalous tails exist also for nonlinear waves.

We consider the semilinear wave equation with the power nonlinearity
\begin{equation}\label{nlw}
\Box \phi = \phi^p\,,\qquad \Box=  \partial_t^2  -\Delta\,,
\end{equation}
in odd $d\geq 5$ spatial dimensions ($p\geq 2$ is an integer). We assume that initial data are
small, smooth, spherically symmetric, and compactly supported
\begin{equation}\label{id}
    \phi(0,r)=\varepsilon f(r),\qquad \partial_t\phi(0,r)=
    \varepsilon g(r)\,.
\end{equation}
It is well known that the corresponding solutions exist globally in time (see, for instance,
\cite{gls}) so there arises a natural question: what is the asymptotic behavior of solutions for
$t\rightarrow \infty$~? In the following we address this question using perturbation theory. Our
starting point is the perturbation expansion
\begin{equation}\label{pert}
    \phi=\varepsilon \phi_0 + \varepsilon^2 \phi_1+\varepsilon^3 \phi_2+...\,,
\end{equation}
where $\varepsilon \phi_0$ satisfies initial data (\ref{id}) and all
higher $\phi_n$ have zero initial data. Substituting this expansion
into equation (\ref{nlw}) we get the iterative scheme
\begin{subequations}
\begin{eqnarray}
  \Box \phi_0 &=& 0\,, \label{f1p}\\
  \Box \phi_{p-1} &=& \phi_0^p\,, \label{f2p}\\
  \Box \phi_{2p-2} &=& p\, \phi_0^{p-1} \phi_{p-1}\,,\qquad \mbox{etc,} \label{f3p}
\end{eqnarray}
\end{subequations}
which can be solved recursively.

 The general
spherically symmetric solution of equation (4a) is given a
superposition of outgoing and ingoing waves
\begin{equation}\label{phi0}
    \phi_0(t,r)=\phi_0^{ret}(t,r)+\phi_0^{adv}(t,r)\,,
\end{equation}
where
\begin{equation}\label{phi0p}
    \phi_0^{ret}(t,r)= \frac{1}{r^{l+1}}\,\sum_{k=0}^{l}  \frac {(2l-k)!} {k!(l-k)!} \frac
{a^{(k)}(u)}{(v-u)^{l-k}}\,, \qquad   \phi_0^{adv}(t,r) =
\frac{1}{r^{l+1}}\,\sum_{k=0} ^{l} (-1)^{k+1} \frac {(2l-k)!}
{k!(l-k)!} \frac {a^{(k)}(v)}{(v-u)^{l-k}}\,,
\end{equation}
and $u=t-r$, $v=t+r$ are the retarded and advanced times, respectively (the superscript in round
brackets denotes the $k$-th derivative). Here and in the following, it is convenient to use the
positive integer index $l$ defined by $d=2l+3$ (recall that we consider only \textit{odd} spatial
dimensions $d\geq 5$).
 Note that for
compactly supported initial data the generating function $a(x)$ can
be chosen to have compact support as well (this condition determines
$a(x)$ uniquely).

 To solve equation (4b) we use
   the Duhamel formula for the solution of the inhomogeneous
   equation $\Box \phi=N(t,r)$ with zero initial data (see, e.g.,
   \cite{ls})
\begin{equation}\label{duh}
    \phi(t,r)= \frac{1}{2 r^{l+1}}
    \int\limits_{0}^{t} d\tau \int\limits_{|t-r-\tau|}^{t+r-\tau} \rho^{l+1} P_l(\mu)
 N(\tau,\rho) d\rho\,,
\end{equation}
where $P_l(\mu)$ are Legendre polynomials of degree $l$ and
$\mu=(r^2+\rho^2-(t-\tau)^2)/2r\rho$ (note that $-1\leq \mu \leq 1$
within the integration range).
Applying this formula to equation (4b) and using null coordinates
$\eta=\tau-\rho$ and $\xi=\tau+\rho$ we obtain
\begin{equation}\label{phi1}
    \phi_{p-1}(t,r)= \frac{1}{2^{l+3} r^{l+1}}
    \int\limits_{|t-r|}^{t+r} d\xi \int\limits_{-\xi}^{t-r} (\xi-\eta)^{l+1} P_l(\mu)
    \phi_0^p(\eta,\xi)  d\eta\,,
\end{equation}
where now $ \mu=(r^2+(\xi-t)(t-\eta))/r(\xi-\eta)$.
If the initial data (\ref{id}) vanish outside a ball of radius $R$,
then for $t>r+R$ we may drop the advanced part of $\phi_0$ and
interchange the order of integration in (\ref{phi1}) to get
\begin{equation}\label{phi12}
    \phi_{p-1}(t,r) = \frac{1}{2^{l+3} r^{l+1}}
    \int\limits_{-\infty}^{\infty} d\eta  \int\limits_{t-r}^{t+r} (\xi-\eta)^{l+1}
    P_l(\mu)  [\phi_0^{ret}(\eta,\xi)]^p\,
    d\xi\,.
\end{equation}
Substituting (\ref{phi0}) into (\ref{phi12}) and using the identity (see the appendix for the derivation)\footnote{As in \cite{bcr} we use the notation $x^{\underline{0}}
:= 1,\,\,x^{\underline{k}} := x \cdot (x-1) \cdot \dots \cdot (x-(k-1)), \quad k>0.$},
\begin{eqnarray}
\int \limits_{t-r}^{t+r} d\xi \, \frac {P_l (\mu)}
{(\xi-\eta)^{n}} &=&  (-1)^l \frac {2(n-2)^{\underline{l}}}
{(2l+1)!!} r^{l+1} \frac {(t-\eta)^{n-l-2}} {[(t-\eta)^2 - r^2]^{n-1}} \, F \left( \left. \begin{array}{c} \frac {l+2-n} {2}, \, \frac {l+3-n} {2}  \\ l + 3/2 \end{array} \right| \left( \frac {r} {t - \eta} \right)^{2} \right)
\nonumber\\
&=&  (-1)^l \frac {2(n-2)^{\underline{l}}}
{(2l+1)!!} \, \frac {r^{l+1}} {t^{l+n}} \left( 1 + (l+n) \frac {\eta} {t} + \mathcal{O}
\left( \frac {1} {t^2} \right) \right)\, ,
\label{master}
\end{eqnarray}
%\begin{equation}
%\label{master} \int \limits_{t-r}^{t+r} d\xi \, \frac {P_l (\mu)}
%{(\xi-\eta)^{n}} =  (-1)^l \frac {2(n-2)^{\underline{l}}}
%{(2l+1)!!}r^{l+1} \times \frac {1} {t^{l+n}} \left( 1 + \mathcal{O}
%\left( \frac {1} {t} \right) \right)\, ,
%\end{equation}
we get
\begin{equation}
\label{tail1} \phi_{p-1} (t,r) = \frac {C(l,p)} {t^{(l+1)p-1}} \left[ \mathcal{I}_l(p,0) +
\mathcal{O} \left( \frac{1}{t} \right) \right] \,,
\end{equation}
where
\begin{equation}
\label{C} C(l,p) := (-1)^l \frac {2^{(l+1)(p-1)-1}}{(2l+1)!!} [(l+1)(p-1)-2]^{\underline{l}}\,,
%[(l+1)(p-1)-2][(l+1)(p-1)-3]...[(l+1)(p-1)-(l+1)]\,,
%((l+1)(p-1)-2)^{\underline{l}}\,,
\end{equation}
and
\begin{equation}
\label{I} \mathcal{I}_l(p,q) := \int \limits_{-\infty}^{+\infty} (a^{(l)}(\eta))^p
(a^{(l+1)}(\eta))^q \, d\eta.
\end{equation}
 The coefficient $\mathcal{I}_l(p,q)$ is
the only trace of initial data. We point out that there is no loss of generality in putting the
coefficient $C(l,p)$ outside the square bracket in (\ref{tail1}) because if $C(l,p)=0$ (which
happens for $p=2$), then  the integrand over $\eta$ in (\ref{phi12}) becomes a total derivative,
hence the whole integral (\ref{phi12}) vanishes for compactly supported initial data.

It is not difficult to verify that generically $\phi_{2p-2}$ and all higher-order iterates also
decay as $1/t^{(l+1)p-1}$, thus $\phi_{p-1}$ gives a good approximation of the full tail provided
that $\varepsilon$ is sufficiently small. More precisely, for fixed $r$ and $t\rightarrow \infty$
we have
\begin{equation}\label{app}
\phi(t,r)\approx {\varepsilon}^p \phi_{p-1}(t,r)\,,
\end{equation}
 up to an error
of order $\mathcal{O}(t^{-[(l+1)p-1]})
\mathcal{O}(\varepsilon^{2p-1})$. We remark that for $l=0$ the
series (\ref{pert}) was proven in \cite{nik} to converge for small
enough $\varepsilon$, thereby  making the approximation (\ref{app})
rigorous. We have not been able to prove an analogous convergence
result for $l\geq 1$, however from the practical point of view the
asymptotic nature of the perturbation series is sufficient in using
the approximation (\ref{app}) to make quantitative physical
predictions.

The advantage of the approach presented above, in contrast to decay estimates in the form of
inequalities, is that it makes easy to identify anomalous tails for which the amplitude of the
leading order term in the perturbation expansion  vanishes. In the case at hand, as mentioned
above, this happens for the quadratic nonlinearity $p=2$ since from (\ref{C})  the coefficient
$C(l,2)$ vanishes for any $l\geq 1$. This implies that there is no tail in the first order or,
put differently, the system of equations (4a) and (4b) satisfies Huygens' principle (note that
 this is not true  in
 three spatial dimensions, i.e. for $l=0$, \cite{nik}).

In order to determine
the tail for the quadratic nonlinearity we need to go to the second
order.
Applying the Duhamel formula (\ref{duh}) to equation (4c) we obtain
\begin{equation}\label{phi2}
    \phi_2(t,r)= \frac{1}{2^{l+2} r^{l+1}}
    \int\limits_{|t-r|}^{t+r} d\xi \int\limits_{-\xi}^{t-r} (\xi-\eta)^{l+1} P_l(\mu)
    \phi_0(\eta,\xi)\phi_1(\eta,\xi)  d\eta\,.
\end{equation}
To compute the asymptotic behavior of this expression near timelike
infinity  we need to find first the asymptotic expansion of
$\phi_1(t,r)$ near null infinity ($u=const$ and $v\rightarrow
\infty$). Substituting (\ref{phi0p}) into (4b) we get
\begin{equation}\label{phi1nul}
 \phi_1(u,v) =
 \mbox{free part}+\frac{h(u)}{(v-u)^{2l+1}}+\mathcal{O}(1/v^{2l+2})\,,\qquad
 h(u) = - \frac{2^{2l}}{l}\int_{-\infty}^u [a^{(l)}(x)]^2 dx\,.
\end{equation}
Plugging  (\ref{phi1nul}) and (\ref{phi0p})  into (\ref{phi2}),
interchanging the order of integration, and expanding in powers of
$1/t$, we get the leading order asymptotic behavior at timelike
infinity
\begin{equation}\label{inner2}
\phi_2(t,r) = (-1)^l \frac{2^{3l}}{2l(2l+1)} \, \frac{1}{t^{3l+1}} \, \left[
\mathcal{I}_{l-1}(1,2) + \mathcal{O} \left( \frac{1}{t} \right) \right]\,.
\end{equation}
Thus, for $p=2$ the approximation (\ref{app}) should be replaced by
\begin{equation}\label{app2}
\phi(t,r)\approx {\varepsilon}^3 \phi_2(t,r)+ \mathcal{O}(\varepsilon^{4})\,,
\end{equation}
where $\phi_2$ is given by (\ref{inner2}).

Since quadratic nonlinearities are common in nonlinear perturbation analysis, the anomalous tail
(\ref{app2}) appears frequently in applications. We emphasize that these applications are not
restricted to higher dimensions and might be physically relevant because some important equations
in physics (we mean, in four dimensions) are equivalent to scalar wave equations in higher
dimensions (actually, we first came across this phenomenon while studying the Yang-Mills
equations in four dimensions \cite{ym}).

The analysis presented above can be readily generalized to incorporate nonlinearities involving
derivatives, for example equations of the form
\begin{equation}\label{grad}
\Box \phi=\phi^p(\alpha \partial_t\phi + \beta \partial_r\phi)^q\,, \end{equation} where integers
$p$ and $q$ satisfy $p+q\geq 2$, and $\alpha,\beta$ are constants. For this equation, proceeding
along the same lines as in the derivation of the tail (\ref{tail1}), we obtain
\begin{equation}
\label{tail2} \phi_{p+q-1} (t,r) = (\alpha - \beta)^q \frac {\,C(l,p+q)} {t^{(l+1)(p+q)-1}}
\left[ \mathcal{I}_l(p,q) + \mathcal{O} \left( \frac{1}{t} \right) \right]\,.
\end{equation}
There are several special cases when the tail (\ref{tail2}) vanishes and the decay is faster:
\begin{itemize}
\item
$q=0$ and $p=2$. This case has been discussed above (see(\ref{inner2})).
\item
$q=1$ and $p\geq 1$.  In this case the coefficient $\mathcal{I}_l(p,1)$ vanishes (since the
integrand in (\ref{I}) is the total derivative and $a(x)$ is compactly supported) and, instead of
(\ref{tail2}), we have
\begin{equation}
\label{q1p11st} \phi_p(t,r) =  \frac{D(l,p)}{t^{(l+1)(p+1)}} \left[ \mathcal{I}_l(p+1,0) +
\mathcal{O} \left( \frac{1}{t} \right) \right]\,,
\end{equation}
where
\begin{equation}\label{btilde}
D(l,p) = (-1)^l \frac {2^{(l+1)p - 1}} {(2l+1)!!} (l+1) [(l+1)p-1]^{\underline{l}} \left[
(\beta-\alpha) \frac {p-1} {p+1} \frac{(l+1)(p+1)-1} {(l+1)p-1} - 2 \beta \right]\,.
\end{equation}
Note that for $p=1$ and $\beta=0$ the coefficient $D(l,p)$ vanishes. In this case there is no
tail at the first order whatsoever, while at the second order, in analogy to (\ref{inner2}), we
obtain
\begin{equation}\label{}
\phi_2(t,r) = (-1)^l \alpha^2 2^{3l-2} \frac{3l+1}{2l(2l+1)} \, \frac {1}{t^{3l+2}}
 \left[ \mathcal{I}_l(3,0) + \mathcal{O} \left( \frac{1}{t} \right) \right]\,.
\end{equation}
\item
$q=2$ and $p=0$. In this case $C(l,p+q)=0$ and, instead of (\ref{tail2}), we have
\begin{equation}
\label{q2p01st} \phi_1(t,r) = (-1)^l \alpha \beta \frac {2^{l+2} l!} {(2l+1)!!} (l+1)^3 \,
\frac{1}{t^{2l+3}} \left[ \mathcal{I}_l(2,0) + \mathcal{O} \left( \frac{1}{t} \right) \right]\,,
\end{equation}
thus for $\alpha\beta=0$ there is no first order tail, in analogy to the case $q=0$ and $p=2$. At the second order we get
\begin{equation}
\label{q2p02nd} \phi_2(t,r) = (-1)^{l+1} (\alpha-\beta)^4 \frac{2^{3l}}{2l(2l+1)} \,
\frac{1}{t^{3l+1}} \, \left[ \mathcal{I}_l(0,3) + \mathcal{O} \left( \frac{1}{t} \right)
\right]\,.
\end{equation}
We note that for $l=1$ and $0\neq \alpha \neq \beta \neq 0$ this case is exceptional in the sense
that the first-order tail decays faster then the second-order tail. This is in fact a peculiar
property of the nonlinearity of the form $\partial_t\phi \partial_r\phi$ in 5+1 dimensions, as
$\left(\alpha\partial_t\phi + \beta\partial_r\phi\right)^2 =
\alpha^2\left(\partial_t\phi\right)^2 + 2 \alpha\beta\partial_t\phi \partial_r\phi +
\beta^2\left(\partial_r\phi\right)^2$.
\\
Note also that from (\ref{q2p01st}) and (\ref{q2p02nd}) the nonlinearities
$\left(\partial_t\phi\right)^2$ and $\left(\partial_r\phi\right)^2$ produce exactly the same
tail, in agreement with the well-known fact that the equation $ \Box
\phi=\left(\partial_t\phi\right)^2 - \left(\partial_r\phi\right)^2$ is Huygensian.
\item
$q \geq 1$ and $\alpha = \beta \neq 0$. In this case we have
\begin{equation}
\label{alpha=beta} \phi_{p+q-1} (t,r) = \frac {E(l,p,q)} {t^{(l+1)(p+q)+q-1}} \, \left[
\mathcal{I}_l(p+q,0) + \mathcal{O} \left( \frac{1}{t} \right) \right]\,,
\end{equation}
where
\begin{equation}
\label{E} E(l,p,q) = (-1)^{l+q} \alpha^q \frac {2^{(l+1)(p+q-1)+q-1}}{(2l+1)!!} (l+1)^q
((l+1)(p+q-1)+q-2)^{\underline{l}}\,.
\end{equation}
The formula (\ref{alpha=beta}) reduces to (\ref{q1p11st}) if $q=1$ and $p=1$, and to
(\ref{q2p01st}) if $q=2$ and $p=0$.
\end{itemize}
Finally, we wish to remark that all the above analytic predictions have been scrupulously
verified numerically.

\vskip 0.3cm \noindent \textbf{Acknowledgment:} We thank Nikodem Szpak for discussions. This
research was supported in part by the grants 1PO3B01229, 189/6. PR UE/2007/7, and FWF P19126-N16.

\appendix*

%%%%%%%%%%%%%%%%%%%%%%%%%%%%%%%%%%%%%%%%%%%%%%%%%%%%%%%%%%%%%%%%%%%%%%%%%%%%%%%%%%%%%%%%%%%%
%%%%%%%%%%%%%%%%%%%%%%%%%%%%%%%%%%%%%%%%%%%%%%%%%%%%%%%%%%%%%%%%%%%%%%%%%%%%%%%%%%%%%%%%%%%%
%%%%%%%%%%%%%%%%%%%%%%%%%%%%%%%%%%%%%%%%%%%%%%%%%%%%%%%%%%%%%%%%%%%%%%%%%%%%%%%%%%%%%%%%%%%%
\section{}
%%%%%%%%%%%%%%%%%%%%%%%%%%%%%%%%%%%%%%%%%%%%%%%%%%%%%%%%%%%%%%%%%%%%%%%%%%%%%%%%%%%%%%%%%%%%
%%%%%%%%%%%%%%%%%%%%%%%%%%%%%%%%%%%%%%%%%%%%%%%%%%%%%%%%%%%%%%%%%%%%%%%%%%%%%%%%%%%%%%%%%%%%
%%%%%%%%%%%%%%%%%%%%%%%%%%%%%%%%%%%%%%%%%%%%%%%%%%%%%%%%%%%%%%%%%%%%%%%%%%%%%%%%%%%%%%%%%%%%
\noindent Here we derive the identity (\ref{master}). Changing the integration variable from
$\xi$ to $\mu$ we get
\begin{equation}
\int \limits_{t-r}^{t+r} d\xi \, \frac {P_l (\mu)} {(\xi-\eta)^{n}} = \frac {r (t-\eta)^{n-2}} {[(t-\eta)^2 - r^2]^{n-1}} \int \limits_{-1}^{+1} d\mu \, P_l (\mu) \left( 1 - \frac {r\mu} {t-\eta} \right)^{n-2} \,.
\end{equation}
Expanding the power and integrating using the identity \cite{wolfram}
\begin{equation}
\label{mu-to-k} \mu^k = \sum_{l=k,k-2,k-4,\dots} \frac {(2l+1) k!}
{2^{(k-l)/2} \left( \frac {k-l} {2} \right)! (k+l+1)!} P_l (\mu)\,,
\end{equation}
we obtain
\begin{equation}
\int \limits_{t-r}^{t+r} d\xi \, \frac {P_l (\mu)} {(\xi-\eta)^{n}} = (-1)^l r^{l+1} \frac {(t-\eta)^{n-l-2}} {[(t-\eta)^2 - r^2]^{n-1}} \sum _{0 \leq m}
2^{1-m} \left( \begin{array}{c} n-2 \\ l + 2m \end{array} \right) \frac {(l + 2m)!} {m! (2l + 2m + 1)!!}
\left( \frac {r} {t - \eta} \right)^{2m} \,.
\end{equation}
Expressing the sum in terms of the hypergeometric function we get (\ref{master}).


\begin{thebibliography}{10}

%\bibitem{g} P. G\"unther, \textit{Huygens' principle and hyperbolic
%differential equations} (Academic Press, San Diego, 1988).

\bibitem{bcr} P. Bizo\'n, T. Chmaj, and A. Rostworowski,
 Phys. Rev. \textbf{D76}, 124035 (2007)

%\bibitem{k} J. G. Kingston, Quart. Appl. Math. \textbf{46}, 775 (1988).

\bibitem{gls} V. Georgiev, H. Lindblad and C. D. Sogge, Amer. J. Math. \textbf{119}, 1291 (1997).

\bibitem{ls} H. Lindblad and C. D. Sogge, Amer. J. Math. \textbf{118}, 1047 (1996).

\bibitem{nik} N. Szpak, P. Bizo\'n, T. Chmaj, and A. Rostworowski, math-ph/0712.0493

\bibitem{ym} P. Bizo\'n, T. Chmaj, and A. Rostworowski,
Class. Quantum. Grav. \textbf{24}, F55 (2007).

\bibitem{wolfram} http://mathworld.wolfram.com/

\end{thebibliography}
\end{document}